%% file: main.tex
\documentclass[conference]{IEEEtran}
\usepackage{cite}
\usepackage{algorithmic}
\usepackage{textcomp}
\def\BibTeX{{\rm B\kern-.05em{\sc i\kern-.025em b}\kern-.08em
    T\kern-.1667em\lower.7ex\hbox{E}\kern-.125emX}}

\usepackage{mathptmx} 

\newcommand{\ignore}[1]{}
\usepackage{fancyhdr}
\usepackage[normalem]{ulem}
\usepackage[hyphens]{url}
\usepackage{microtype}
\usepackage[frozencache,cachedir=_minted-output]{minted}
\usepackage{graphicx}
\usepackage[export]{adjustbox}
\usepackage{subfig}

\usepackage[ruled,vlined,linesnumbered]{algorithm2e}

\usepackage{amsmath}
\usepackage{amssymb}
\usepackage{amsfonts}

\usepackage{enumitem}
\usepackage[binary-units]{siunitx}
\usepackage{soul}
\usepackage[table]{xcolor}
\definecolor{LinkOrange}{HTML}{F9CB9C}
\definecolor{LinkBlue}{HTML}{A4C2F4}
\definecolor{LinkGreen}{HTML}{B6D7A8}

\usepackage{booktabs}
\usepackage{multirow}
\usepackage{array}
\newcolumntype{L}[1]{>{\raggedright\let\newline\\\arraybackslash\hspace{0pt}}m{#1}}
\newcolumntype{C}[1]{>{\centering\let\newline\\\arraybackslash\hspace{0pt}}m{#1}}
\newcolumntype{R}[1]{>{\raggedleft\let\newline\\\arraybackslash\hspace{0pt}}m{#1}}
\usepackage[oldenum]{paralist}
\usepackage{dblfloatfix} 
\usepackage[bookmarks=true,breaklinks=false,colorlinks,linkcolor=black,citecolor=blue,urlcolor=black]{hyperref}

\newcommand{\update}[1]{{\color{black}{#1}}}
\newcommand{\marker}[1]{(\textbf{#1})}

\newcommand{\VL}{\textit{VL}\xspace}
\newcommand{\PE}{\textit{PE}\xspace}
\newcommand{\VLong}{Virtual-Link\xspace}
\newcommand{\SQI}{\textit{SQI}\xspace}
\newcommand{\PA}{\textit{PA}\xspace}
\newcommand{\VA}{\textit{VA}\xspace}
\newcommand{\VLRD}{\textit{VLRD}\xspace}

\newcommand{\MN}{\textit{M:N}\xspace}
\newcommand{\CAS}{\textit{CAS}\xspace}
\newcommand{\GEM}{\textit{gem5}\xspace}

\newcommand{\SOC}{\textit{SoC}\xspace}
\newcommand{\ZMQ}{\textit{ZMQ}\xspace}
\newcommand{\boost}{\textit{BLFQ}\xspace} 

\newcommand{\ISA}{\textit{ISA}\xspace}
\newcommand{\IPC}{\textit{IPC}\xspace}
\newcommand{\LONE}{\textit{L1D}\xspace}
\newcommand{\LTWO}{\textit{L2}\xspace}

\newcommand{\SELECT}{\texttt{vl\_select}\xspace} 
\newcommand{\PUSH}{\texttt{vl\_push}\xspace} 
\newcommand{\FETCH}{\texttt{vl\_fetch}\xspace}
\newcommand{\linkTab}{\textit{linkTab}\xspace}
\newcommand{\consBuf}{\textit{consBuf}\xspace}
\newcommand{\prodBuf}{\textit{prodBuf}\xspace}
\newcommand{\prodHead}{\texttt{prodHead}\xspace}

\newcommand{\consHead}{\texttt{consHead}\xspace}
\newcommand{\consTail}{\texttt{consTail}\xspace}
\newcommand{\consTgt}{\texttt{consTgt}\xspace}
\newcommand{\nextIn}{\texttt{nextIn}\xspace}

\newcommand{\CIHR}{\textit{CIHR}\xspace}
\newcommand{\CITR}{\textit{CITR}\xspace}
\newcommand{\CIFR}{\textit{CIFR}\xspace}
\newcommand{\PIHR}{\textit{PIHR}\xspace}

\newcommand{\LL}{\textit{LL}\xspace}
\newcommand{\VQ}{\textit{VQ}\xspace}
\newcommand{\pingpong}{\textit{ping-pong}\xspace}
\newcommand{\incast}{\textit{incast}\xspace}
\newcommand{\halo}{\textit{halo}\xspace}
\newcommand{\sweep}{\textit{sweep}\xspace}

\newcommand{\fir}{\textit{FIR}\xspace}
\newcommand{\bitonic}{\textit{bitonic}\xspace}
\newcommand{\stream}{\textit{STREAM}\xspace}

\newcommand{\pipeline}{\textit{pipeline}\xspace}

\newcommand{\RT}{\texttt{Rt}\xspace} 
\newcommand{\RS}{\texttt{Rs}\xspace} 
\begin{document}
\bstctlcite{IEEEexample:BSTcontrol}
\title{\VLong: A Scalable Multi-Producer, Multi-Consumer Message Queue Architecture for Cross-Core Communication}
\author{
\IEEEauthorblockN{Qinzhe Wu\textsuperscript{\dag}}
\IEEEauthorblockA{qw2699@utexas.edu}
\and

\IEEEauthorblockN{Jonathan Beard\textsuperscript{*}}
\IEEEauthorblockA{jonathan.beard@arm.com}
\and

\IEEEauthorblockN{Ashen Ekanayake\textsuperscript{\dag}}
\IEEEauthorblockA{ashen.ekanayake@utexas.edu }
\and

\IEEEauthorblockN{Andreas Gerstlauer\textsuperscript{\dag}}
\IEEEauthorblockA{gerstl@ece.utexas.edu}
\and

\IEEEauthorblockN{Lizy K. John\textsuperscript{\dag}}
\IEEEauthorblockA{ljohn@ece.utexas.edu}
\and

\IEEEauthorblockA{\hspace{150pt}\textit{\textsuperscript{\dag}The University of Texas at Austin, \textsuperscript{*}Arm Inc.}\\}
}

\maketitle
\pagestyle{plain}

\begin{abstract}
\input{abstract}
\end{abstract}

\vspace{-2mm}
\section{Introduction}\label{sec:intro}
\input{Introduction}

\section{Problem Description and Motivation}\label{sec:motiv}
\input{Motivation}

\section{Design}\label{sec:design}
\input{Design}

\section{Evaluation}\label{sec:eval}
\input{Evaluation}

\section{Related Work}\label{sec:related}
\input{Related}

\section{Conclusion}\label{sec:conclu}
\input{Conclusion}


\bibliographystyle{IEEEtran}
\bibliography{ref}

\end{document}

%% file: abstract.tex
Cross-core communication is increasingly a bottleneck as the number of 
processing elements increase per system-on-chip. Typical hardware 
solutions to cross-core communication are often inflexible; while 
software solutions are flexible, they have performance scaling limitations.
A key problem, as we will show, is that of shared state in  
software-based message queue mechanisms. This paper proposes 
\VLong (\VL), a novel light-weight communication mechanism with hardware 
support to facilitate \MN lock-free data movement.
\VL reduces the amount of coherent shared state, 
which is a bottleneck for many approaches, to zero.
\VL provides further latency benefit by keeping data on the fast 
path (i.e., within the on-chip interconnect). \VL enables directed 
\textit{cache-injection} (\textit{stashing}) between {\PE}s on the 
coherence bus, reducing the latency for core-to-core communication. 
\VL is particularly effective for fine-grain tasks on streaming data.  
Evaluation on a full system simulator with 
$7$
benchmarks shows that \VL achieves a
$2.09\times$
speedup over
state-of-the-art software-based communication mechanisms,
while reducing memory traffic by $61\%$. 

%% file: Introduction.tex
Frequency scaling~\cite{dennard1974design} is no longer a practical 
option for year-over-year performance increase.
With the impending end of lithography 
scaling~\cite{kish2002end},
a world searching for performance is left 
seeking more radical solutions~\cite{vetter2018extreme}. 
Some architects are building up 
and out, others are
searching for domain specific acceleration, or even 
re-configurable accelerators.
Regardless of the combination or modality chosen 
to continue the march towards
increasing performance, a key bottleneck for both efficiency and
performance remains: communication cost~\cite{kleen2009linux}. 
In order for two or more threads of execution 
to work together as a multi-threaded program, they must
be able to message each other, i.e., to 
communicate~\cite{dijkstra1965solution,lamport1979make}. 
The cost of communication bounds the overall parallelism that 
can be extracted~\cite{arvind1988assessing}.
Communication is required for both initiating new parallel work
and for data distribution.
Extant architectures 
are pushing 128-cores per \SOC~\cite{ampere}, yet with current communication/synchronization mechanisms, they are hard to 
fully utilize for a single application. 
Synchronization overheads mount as the number of threads increase, a 
fact that has certainly been noticed~\cite{pasetto2012performance,akkan2013hpc}. 
Depending on the amount of ``compute'' in each parallel kernel per ``firing'', the benefits of parallelization may disappear due to this overhead.

Many solutions for core-to-core 
communication exist, with varying flexibility and hardware support.
Hardware solutions abound,
from direct register transfers~\cite{grafe1989epsilon,papadopoulos1990monsoon,noakes1993j} to active message solutions~\cite{TILE64_ISSCC08,chen2007cell,qoriq2012primer}; these are generally fast but inflexible.
Flexible software solutions range
from lock-based synchronization over a critical section to
the doubly linked-list (\LL) formulations of lock-free queues~\cite{ladan2004optimistic} (see \S~\ref{sec:related}). 
Software queueing libraries often utilize 
convenient synchronization primitives, e.g.
locks.
Locks can take many forms but share one drawback, that of
scalability~\cite{michel2018packet}.
Well-known software solutions
include the Boost Lock-Free Queue (\boost~\cite{boost}) and
ZeroMQ (\ZMQ~\cite{hintjens2010zeromq}). 
\begin{figure}
    \centering
    \includegraphics[
        page=1,
        trim=0mm 0mm 0mm 0mm,
        clip=true,
        natwidth=6.10in,
        natheight=2.86in,
        width=\linewidth
    ]{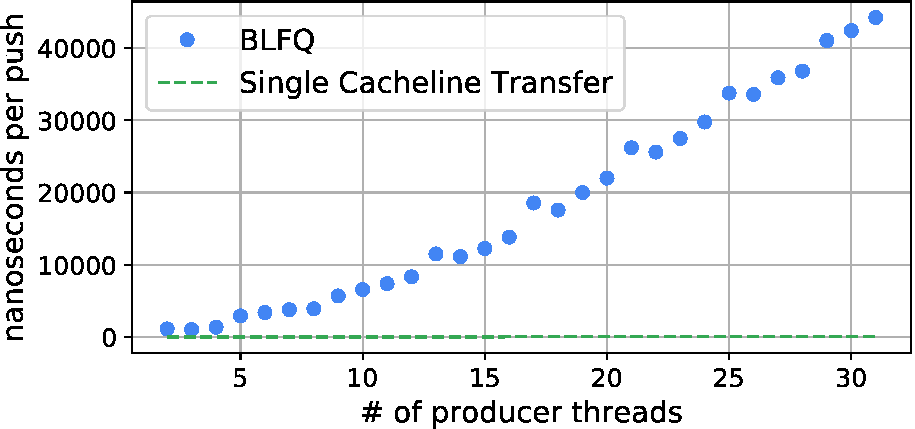}
    \caption{Scaling of a Boost lock-free queue (\boost), varying the count 
    of producers sending data to
    a consumer. The dashed line
    shows the latency observed for 
    transmission of a single cache line between cores without synchronization overheads.}
    \vspace{-5mm}
    \label{fig:boostscaling}
\end{figure}
Figure~\ref{fig:boostscaling} shows a comparison of the \boost
when
scaling the number of producers blue data points). 
If unsynchronized,  a cache line can be transported
between processing elements (\textit{PE}) in \SI{\sim 22}{\nano\second}
to \SI{\sim 34}{\nano\second} green dashed lines),
but this is not achieved if that 
transfer must be synchronized/coordinated. Time per push rises quickly as the number of threads increases,
far and above the dashed line.

This paper introduces \VLong (\VL) 
as a solution to close this communication overhead gap.
\VL is nearly as performant as many hardware-only solutions, 
while being as flexible as the most modern software queue.
Instead of having threads access the shared queue state variables (i.e., head, tail, or lock) atomically,
\VL provides configurable hardware support for \MN communication,
providing both data transfer and synchronization.
Unlike other hardware queue architectures,
\VL reuses the existing cache coherence network and delivers
a virtualized channel as if there were a direct link (or route) between two arbitrary {\PE}s.
\VL facilitates efficient synchronized data movement between \MN 
producers and consumers with several benefits:
\begin{inparaenum}[(i)]
\item the number of sharers on synchronization primitives is reduced to zero, eliminating a
primary bottleneck of traditional lock-free queues,
\item memory spills, snoops, and invalidations are reduced, 
\item data stays on the fast path (inside the interconnect) 
a majority of the time.
\end{inparaenum}

\noindent
The contributions of this work are:
\begin{enumerate}
    \item
    We present a characterization of communication bottlenecks existing in modern software queues based on measurements.
    \item We propose \VLong,
    a hardware/software solution to eliminate the overhead of synchronization,
    achieving efficient synchronization between producers and consumers.
    \item We perform evaluation on
    7
    benchmarks.
    \VL synchronization significantly improve the performance by
    2.09$\times$
    on average over conventional synchronization.
\end{enumerate}

\noindent
In the next sections, we describe how extant solutions
(both lock-based and lock-free)
scale on modern systems,
identifying the synchronization issues to solve (\S~\ref{sec:motiv}),
then elaborate the design of \VL (\S~\ref{sec:design}) based on this
defined problem space, and present the implementation as well 
as evaluation (\S~\ref{sec:eval}).
At the end, we compare \VL with 
related architecture and software solutions (\S~\ref{sec:related}),
and lastly draw conclusions (\S~\ref{sec:conclu}).

%% file: Motivation.tex

\begin{figure}[btp]
    \vspace{-3mm}
    \centering
    \includegraphics[
        page=1,
        clip=true,
        natwidth=4.81in,
        natheight=2.46in,
        width=2.7in
    ]{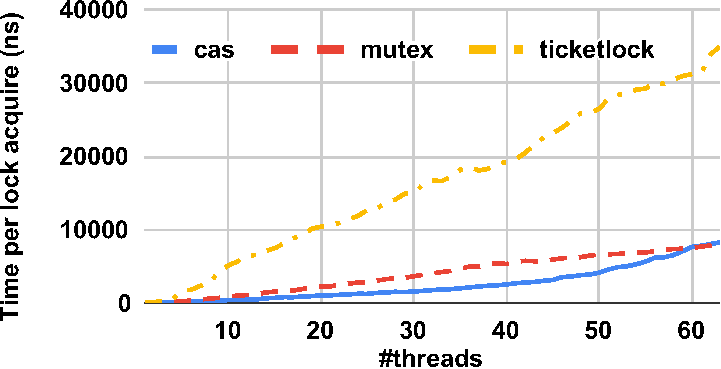}
    \caption{Comparison of execution time in \SI{}{\nano\second} of 
    three different communications mechanisms using the {\it{ lockhammer }}
    benchmark~\cite{lockhammer} on Platform 1 from Table~\ref{table:hwplatform}.
    Regardless of synchronization mechanism, by the time $14$ cores
    are contending for the lock, \SI{\sim 1000}{\nano\second} consumed 
    per lock.}
    \vspace{-0.2in}
    \label{fig:lockgraph}
\end{figure}

There are two integral parts in sending a message:
\textit{atomicity} and \textit{condition synchronization}~\cite{scott2013shared}.
There are various means to achieve the latter
(loosely ordered
from complex to simple, and by no means intended to be complete): 
Compare-And-Swap (\CAS), spin-locks, 
load-linked store-conditional, ticket-locks, and so on.
Building on each of these mechanisms, programmers can guarantee exclusivity 
of access to a critical section that constitutes a region for data communication.  
Figure~\ref{fig:lockgraph} shows 
a sweep of a \CAS-based lock, a ticket-lock, and a standard 
spin-lock on a Platform 1 from Table~\ref{table:hwplatform} using the open-source
\textit{lockhammer}~\cite{lockhammer} benchmark.
Even for a \CAS-based lock, 
after a relatively small number of threads,
the overhead to acquire a lock becomes high enough to ensure
programmers who want to write efficient programs stick to
extremely coarse-grained parallel kernels to amortize synchronization costs.

\begin{figure}[!btp]
    \vspace{-3mm}
    \centering
    \includegraphics[width=2.75in,height=1.12in]{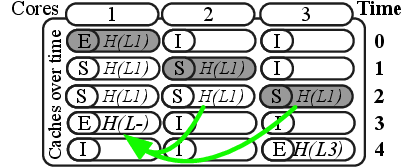}
    \caption{
    Behavior of a single ``lock'' pointer on three cores.
    }
    \vspace{-5mm}
    \label{fig:lockbehavior}
\end{figure}

\begin{figure}[bhtp]
    \vspace{-0.1in}
    \centering
    \includegraphics[natwidth=6.10in,natheight=2.47in,width=0.89\linewidth]{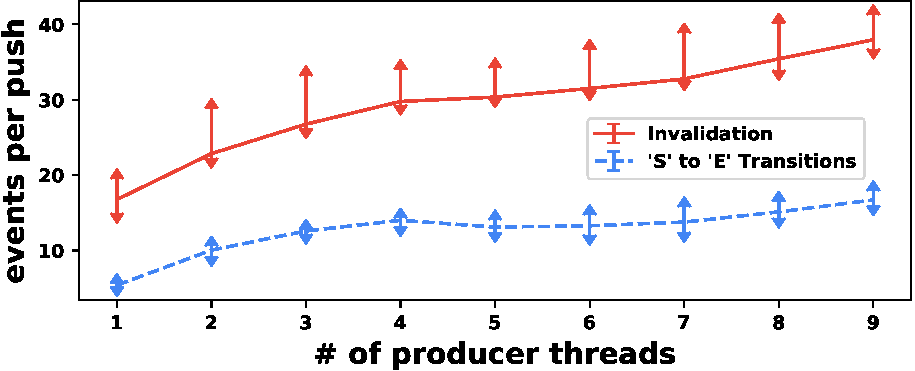}
    \caption{
    Cache events measured per Boost Lock-free Queue push.
    The red line (top) represents the number of 
    invalidations,
    the blue line (bottom) the 
    \textit{shared} to \textit{exclusive} transitions.
    } 
    \vspace{-0.1in}
    \label{fig:boostPC}
\end{figure}

Figure~\ref{fig:lockbehavior} depicts the behavior of a single ``lock'' variable being operated on atomically, as is the case with \CAS, by three cores. For each instance of time, \marker{S} represents a 
``shared'' cache state, \marker{E}
an ``exclusive'' state, and \marker{I} indicates that the 
cache line is 
invalid.
The \marker{L-\textit{X}} represents a lock while the \textit{X}
indicates which core owns the lock. The arrows from Time 2 to Time 3
represent the invalidate-acknowledge traffic that must occur before Core
1 can release the lock.
\textbf{It is this traffic, and therefore the number of
sharers that bound synchronization performance.}
Figure~\ref{fig:boostPC} shows empirical 
performance counter measurements from Platform 2 from Table~\ref{table:hwplatform} (chosen for counter availability) demonstrating that the 
number of invalidation events and \textit{shared} to \textit{exclusive}
coherence state transitions increase proportionally with the number
of sharers (in this case the number of producer threads). 
With a significant number of contending sharers, cores, and large
interconnect, the time to perform a \CAS operation can be 
sizeable.
Thus, while data movement itself between cores
in a coherence network is quite fast~\cite{martin2012chip},
updates to widely shared variables
(e.g. a queue head pointer) can take a significant 
amount of time. 
Based on this observation,
\VL adds hardware support to manage the shared queue state,
and assign unique endpoints for each producer or consumer thread
to operate free of contention.

\textbf{An efficient queue mechanism needs back-pressure:}
Real systems experience some form of transient rate
mismatch between otherwise rate matched producer-consumer 
pairs
causing ``bursty'' queue occupancy to be 
observed~\cite{harchol2013performance}.
As such, any solution
must provide a low-overhead mechanism to accommodate this behavior.
Providing back-pressure when a queue is full is necessary to
prevent buffer spillage to memory or overwriting of contents. 
Hence, we incorporate in \VL a low-overhead mechanism to produce back-pressure
as needed, ensuring data can stay within 
the cache coherence interconnect (fast-path) when possible. Without the back-pressure,
programmers must increase buffer size to accommodate ``bursty'' behavior,
increasing the probability of access to main
memory (see Little's
Law~\cite{bertsimas1995distributional}).
We will show that \VL also reduces main memory (DRAM)
access, reducing communication latency and increasing efficiency (DRAM access is $\gtrapprox100\times$ more expensive in terms 
of energy than SRAM~\cite{jiang2020cimat}). 
Likewise, when arrival rates are greater than 
consumer service rates, back-pressure enables software to 
perform adjustments such as changing the \PE configuration,
or throttling compute kernels. 

\textbf{Smart cache line injection:} 
Traditional software message queues typically load data (or shared state variables) on 
demand.
At best, these queues rely on 
prefetching to ensure the data is near vs. far. 
A design feature driving \textit{VL}'s mechanism is the ability 
to target and stash data to endpoints directly, e.g.
the local private L1 data cache.
This should result in a latency advantage ($\sim2\times$
faster~\cite{leon2011cache}). 
Some injection mechanisms must know the target core in order to
target the last level private cache (preferable~\cite{flynn1997amba}), 
other mechanisms that simply target the system-level 
cache~\cite{farshin2020reexamining} do not require this. 
When building systems that rely on knowing who the target
physical cores are ahead of time,
yet another layer of 
synchronization and complexity is added,
e.g. if thread migration is allowed, every producer would
need to look-up consumer targets in a shared table, likely 
demonstrating the same scaling shown in Figure~\ref{fig:lockgraph}.
Additionally, exposing the physical core-id can be a security
risk~\cite{huang2019comparative}, e.g. a virtual-CPU typically 
has no idea what CPU it is actually running
on~\cite{rao2013optimizing}.
Our \VL design must not require the
producers or the consumers to know about each other,
and \VL should allow direct injection
(to transfer data and notify the consumer).

%% file: Design.tex
\begin{figure}
\centering
\includegraphics[height=1.75in,width=2.75in]{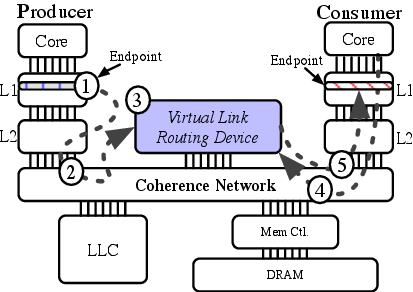}
\caption{
A cache line moves from the producer at \marker{1}, at its own
unique address location, to an indirection layer in hardware at 
\marker{2}. That indirection layer, the 
Routing Device, matches the 
\SQI at \marker{3} based on consumer endpoint demand which is 
registered by \marker{4}. The Routing Device
forwards data to the target consumer buffer on a totally different address
at \marker{5}.
}
\vspace{-0.3in}
\label{fig:highLevel}
\end{figure}

\VLong (\VL) accomplishes the movement of cache lines from producers to consumers by attaching a routing device (\VLRD) to the coherence network as illustrated in Figure~\ref{fig:highLevel}.
\update{The \VLRD is attached to the coherence 
network like a tightly coupled  accelerator or 
system cache slice, from a port on the coherence 
network. 
} 
This \VLRD enables \VL 
to ``link'' unique ``endpoints'' together via a shared queue identifier (\SQI).
Endpoints subscribe to a \SQI to 
form a \MN message channel 
Each \SQI can support \textit{M} producer endpoints and \textit{N} consumer endpoints.
Each unique endpoint for a \SQI maintains its
own local user-space buffer
composed of multiple
coherence granules or cache lines.
Messages from each endpoint are received by the \VLRD at a coherence
granularity, in a lock-free manner.
In abstract,
\textbf{\VL enables a virtual linking of cache lines from each
unique endpoint subscribing to a {\SQI} so that a producer can copy-over data from its own cache lines
directly into a requesting consumer through a single level of indirection.}

\begin{figure}
\centering
\includegraphics[height=1.75in,width=3in]{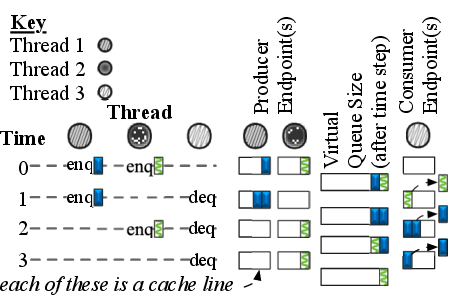}
\caption{Virtual Queue (\VQ) per time step.
2 producer endpoints (Threads 1, 2), 1 consumer endpoint (Thread 3), shares a \SQI.}
\vspace{-0.2in}
\label{fig:qorder}
\end{figure}

Multiple endpoints on a single \SQI come together to form
a Virtual Queue (\VQ).
Figure~\ref{fig:qorder} illustrates
the ordering of operations between two producer endpoints
and a consumer endpoint sharing a 
\SQI. The \VQ size
is shown after each time step.
In
Figure~\ref{fig:qorder}, the cache lines are moved atomically,
that is at time step $2$, the blue producer cache line data appears
to be copied-over atomically (through the interconnect, not main memory) 
to the consumer endpoint buffer.
This copy-over operation leaves the producer cache line zeroed 
and in an exclusive state,
which can be used for subsequent 
enqueue operations.
After the copy-over operation,
the data are shipped to the consumer
to dequeue, also in 
an exclusive state.
At no point does the consumer or producer
access a shared Physical Address (\PA) or Virtual Address (\VA) that could cause coherence traffic (\textit{snoops}).
Instead,
threads check the endpoints owned by themselves and interact with the \VLRD 
for synchronization.
The rest of this section presents the major components of \VL, namely,
the \VLRD, \ISA extensions and system software support. 

\input{routingdevice}


\input{isa}

\input{lib}

\subsection{Enqueue and dequeue}~\label{subsec:control}
\vspace{-.2in}

\begin{figure}[H]
\vspace{-.15in}
\centering
\includegraphics[
clip=true,
natwidth=9.69in,
natheight=2.07in,
width=0.79\linewidth
]{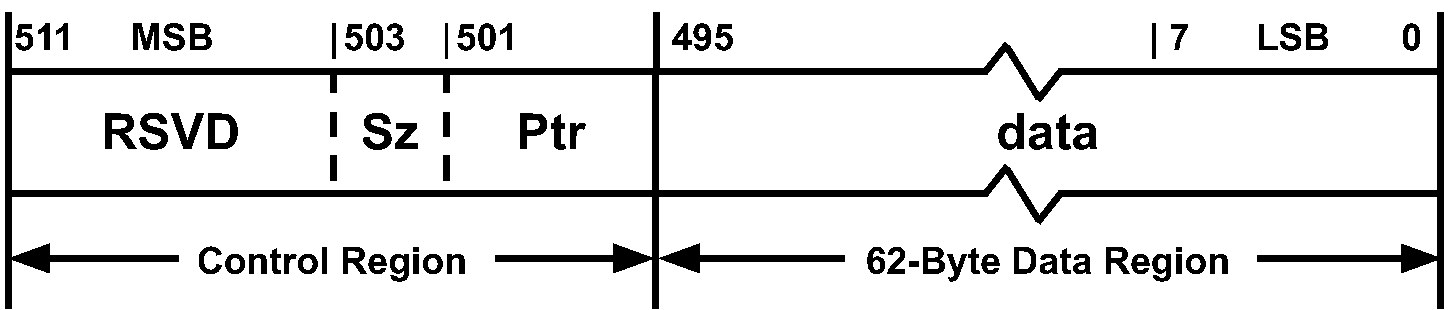}
\caption{Control region and data region in a \SI{64}{\byte} cache line.}
\vspace{-.15in}
\label{fig:cacheline}
\end{figure}

Figure~\ref{fig:qorder} shows the queue order
per single \SQI atomically pushing a \SI{64}{\byte} 
cache line size messages from \MN producer/consumer pairs. 
Messages larger than a cache line can
be incorporated via indirect buffers as pointers. While not 
demonstrated in this paper, it is trivial to incorporate 
an existing indirect buffer format such as VirtIO 1.1~\cite{virtio},
injection could be accelerated in this case by~\cite{RevereAMU_ARM2019}.
To facilitate small message transfer, we embed cache line 
local queue state into the line itself (see Figure~\ref{fig:cacheline}). This consists 
of a \SI{2}{\byte} control region at the Most Significant 
Byte (\textit{MSB}) of each \VL transported cache line. 
the remaining \SI{62}{\byte} are user-data/payload. 
Valid data fills the data region from
higher address towards Least Significant Byte 
(\textit{LSB}). Within the control region, 2b encodes
for size, e.g., byte, half word, word, double word. 
6b encodes a cache line relative offset/head pointer. 
The remaining \SI{1}{\byte} is reserved. 

\noindent\textbf{enqueue:}
With 
respect to Figure~\ref{fig:overallHardware}, the enqueue operation
calls \SELECT at \marker{P1} on an allocated user-space buffer ($Y$). 
The user-space cacheable memory transitions to a ``selected''  state
at \marker{H1} that causes this cache line's \VA to be translated and latched. 
The follow-on \PUSH instruction at \marker{P2} 
causes the cache line at the aforementioned latched \PA from \SELECT ($Y$) to be 
stored to the mapped \VLRD device-memory address ($X$). 
Assuming the conditional store was successful,
the original cache-able user-space memory from $Y$
is owned by the \VLRD. This order of events is necessary to 
prevent a single instruction from requiring two address generations
simultaneously.
If the enqueue succeeds, the 
cache line is zeroed,
otherwise
the return register (see \S~\ref{push}) is set appropriately
so that the programmer can retry pushing the same data
at some future point. 


\noindent\textbf{dequeue:}
Dequeue operations for \VL are essentially operations that 
set a cache line as ``pushable'' while also notifying the 
\VLRD that \SI{64}{\byte} of data is requested at a specific 
cacheable-memory \VA. With respect to 
Figure~\ref{fig:overallHardware}, the dequeue operation 
calls \SELECT at \marker {C2} on an allocated user-space consumer buffer, 
after determining at \marker{C1} that no more data is available (e.g.
by inspecting the control region). Calling \SELECT at \marker{C2} 
sets that \VA and latches the \PA of that line for a follow-on \FETCH instruction 
(\S~\ref{sec:isa}). As described in \S~\ref{sec:isa}, \FETCH 
sets a ``pushable'' flag at \marker{C3} for the cache line addressed by the 
previous \SELECT statement. Following the setting of the ``pushable'' flag,
\FETCH causes the target \PA and core-id to be registered with the \VLRD
at \marker{C4}. That registered \PA is used when data becomes 
available for a given \SQI for a follow-on injection of data 
to the requester at \marker{C5}.

%% file: routingdevice.tex
\begin{figure*}[!t]
    \vspace{-.4in}
    \centering
    \includegraphics[
        trim=0mm 69mm 0mm 0mm,
        natwidth=9.71in,
        natheight=5.56in,
        clip=true,
        width=\linewidth
    ]{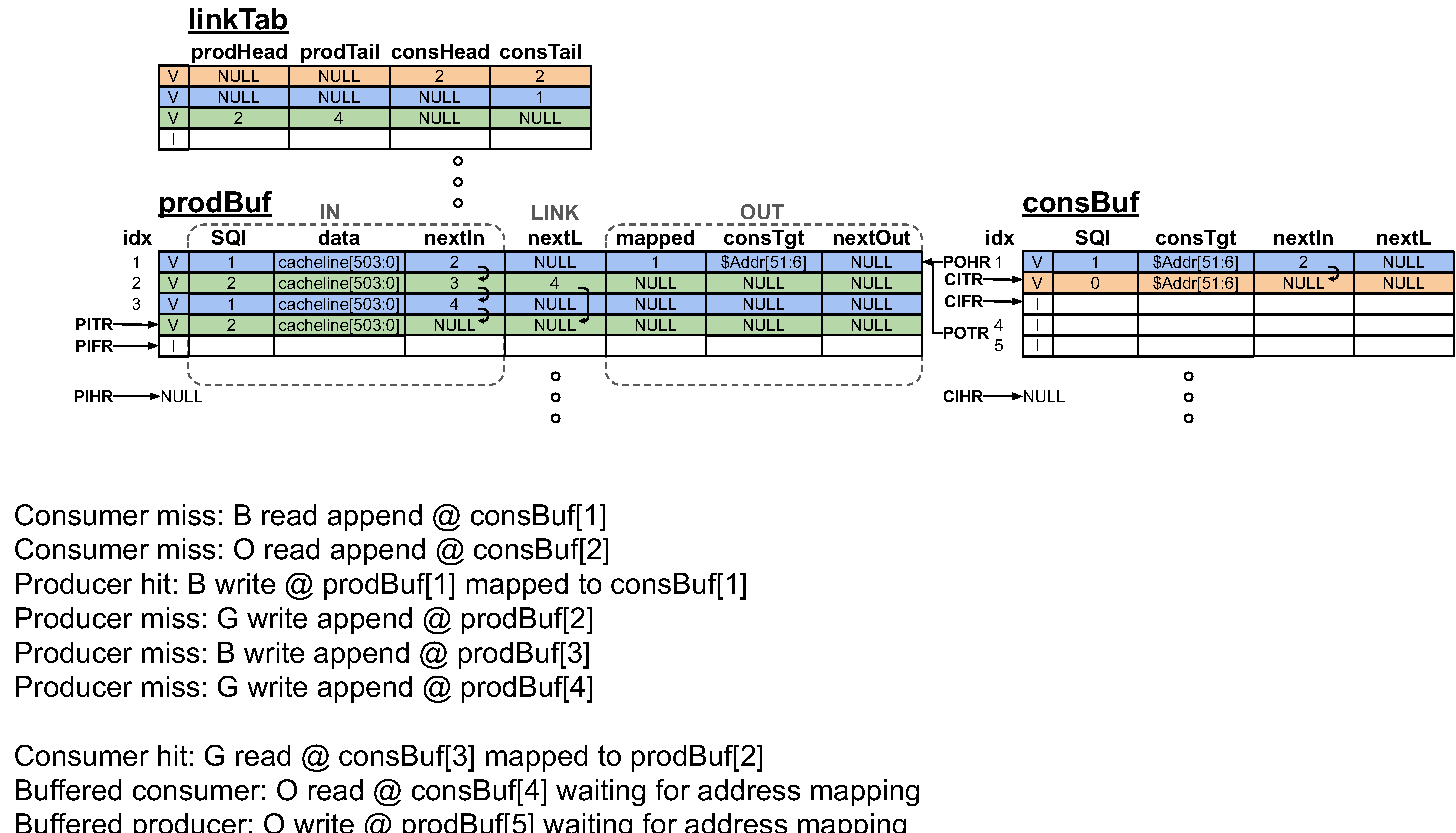}
    \caption{Table and buffer structures in the \VLRD.
    Cells having the same background color belong to the same \SQI.}
    \vspace{-.26in}
    \label{fig:routing_device}
\end{figure*}

\input{routingtablediagram}

\subsection{Routing Device}\label{sec:routing}

The \VLRD is  tasked with matching
incoming messages to a \SQI and stashing those messages
to the subscribed consumers. 
As Figure~\ref{fig:routing_device} shows, 
the \VLRD is largely composed of three structures, the
Link Table (\linkTab), the
Producer Buffer (\prodBuf), and the Consumer Buffer (\consBuf) (some control logic is omitted for brevity).
The \linkTab keeps metadata (i.e., head, tail) for each \SQI,
one per row.
The \prodBuf and \consBuf are shared across multiple \SQI entries, and 
buffer producer data and consumer requests, respectively. Buffer slots are taken in turn and shared by multiple {\SQI}s, therefore these 
structures cannot be used as contiguous FIFOs but instead are 
managed as \update{linked-lists} ({\LL})s.

\noindent\textbf{\underline{linkTab}}:
The head/tail pointers in \linkTab 
each point to the first and last entries in
a hardware-managed inter-leaved \LL data 
structure, which enables hardware to determine
whether there is consumer demand on a specific \SQI
or data available from a producer to send.
The producer head (\prodHead)
is updated
if
the current head is mapped and ready to be sent to a consumer.
For example in Figure~\ref{fig:routing_device} the green row,
\prodHead points to index 2
(Row 2
in \prodBuf).
Once index 2 is mapped with a green consumer request coming later,
\prodHead
is set to
the next green entry (4 in this example).
The \linkTab is addressed by the \SQI field in \prodBuf and \consBuf.

\noindent\textbf{\underline{consBuf}}:
Whenever a consumer request
arrives in the \VLRD,
the port's control logic checks Consumer Input
Free Register (\CIFR) for a free buffer slot in order to 
buffer the consumer request.
A buffer slot is free if the valid bit is unset,
and \CIFR always moves to the next free slot after a slot is taken,
starting over from the first free \consBuf slot again after touching the bottom.
The consumer request
is composed of two parts:
1) the address of the target consumer cache line
(the local user-space buffer of a consumer endpoint)
buffered in \consTgt as shown in Figure~\ref{fig:routing_device};
and 2) the \SQI of the \VQ from which data is requested.
The former is the payload of the incoming packet,
and latter
is encoded in
the device-memory physical address received through the coherence network (details in \S~\ref{subsub:endpoint}).
The \texttt{nextL} field together with the \consHead, \consTail in \linkTab make {\LL}s for {\SQI}s.
As mentioned before,
the slots in \consBuf is not always used in order when multiple {\SQI}s are active.
The \nextIn field
together with Consumer Input Head Register (\CIHR) and Consumer Input Tail Register (\CITR) forming a \LL,
so that \consBuf can track the order to feed the address mapping pipeline.
Address mapping pipeline stages are illustrated in Table~\ref{tab:pipeline} (explained later).
\vspace{-4pt}

\noindent\textbf{\underline{prodBuf}}:
The Producer Buffer has three partitions,
namely, \textbf{IN}, \textbf{LINK}, \textbf{OUT} 
as shown in Figure~\ref{fig:routing_device}.
On cache line 
arrival to the \VLRD, the Producer Input Free Register (\textit{PIFR})
is checked for a free buffer entry.  The Producer Input Head Register (\textit{PIHR}), 
and Producer Input Tail Register (\textit{PITR}) point to the next, and the last
buffered producer push waiting for address mapping, respectively. 
The \textbf{IN} partition plus the \textbf{LINK} partition are very similar to the \consBuf, except that the \texttt{data} field stores the data enqueued
by producers (\S~\ref{subsec:control}).
The \textbf{LINK} partition is a \LL whose head is the oldest entry ready to be sent to a consumer;
the order in which producer data was received is tracked by the \LL;
so data are sent to consumers in the same order.
The \textbf{OUT} partition is for registering mapped entries,
i.e., entries that have been assigned to a consumer target from the process
in Table~\ref{tab:pipeline}.
For example, the first blue entry in the \prodBuf is mapped to \consBuf entry $1$ as indicated by Figure~\ref{fig:routing_device}.
The \consTgt field in the \textbf{OUT} partition stores
the result of address mapping
(i.e., a target consumer cache line address), and \texttt{mapped} field recording an index 
to the mapped
\consBuf
slot.
There are also two registers
associated with this partition, the Producer Output Head Register (\textit{POHR}),
and the Producer Output Tail Register (\textit{POTR}) to track the next, 
and the last entry ready to send out, respectively.
Each of the three partitions 
is a separate 
SRAM block with its own read/write ports,
making each
partition
accessed independently.

\begin{figure*}[!hbt]
\vspace{-3mm}
\begin{minipage}{0.49\textwidth}
\centering
\subfloat[Hardware view]{
\label{fig:overallHardware}
\includegraphics[width=3in,height=2.5in]{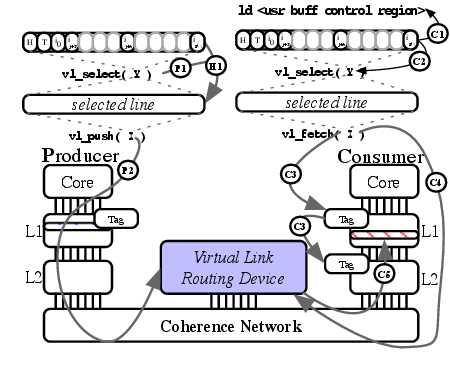}
}
\end{minipage}
\begin{minipage}{0.49\textwidth}
\centering
\subfloat[Software view]{
\label{fig:overallDetailed}
\includegraphics[width=3in,height=2.5in]{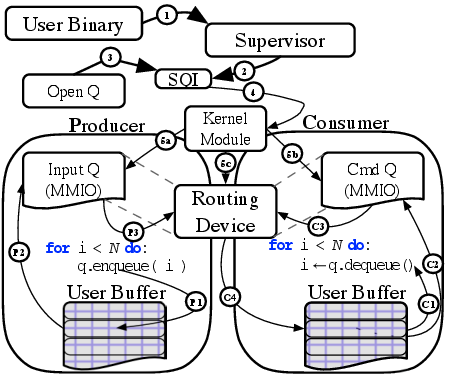}
}
\end{minipage}
\caption{Flow of \VL ISA, hardware, and software 
interaction. (Details in \S~\ref{sec:isa} to
\S~\ref{subsec:control}).
}
\label{fig:HW_SW_views}
\vspace{-4mm}
\end{figure*}

\noindent\textbf{Address mapping}:
A \prodBuf entry with valid data or a \consBuf entry occupied by a consumer request
will go through a 3-stage pipeline illustrated in Table~\ref{tab:pipeline}, to map a producer push with a consumer pull.
At the first stage
the control logic takes \SQI from the ``head entry'' (the entry pointed by either \PIHR or \CIHR) to access the \linkTab and get the head, tail pointers of a corresponding queue.
In 
Stage 2,
a decision is made on whether to map the ``head entry'' to a consumer request or producer data buffered earlier.
For example, in Cycle 1 the first blue consumer request reads blue \prodHead, which is then checked in Cycle 2 Stage 2 in Table~\ref{tab:pipeline}.
The blue request has to append to blue consumer \LL upon a \textit{miss}.
A \textit{hit} occurs in Cycle 4 Stage 2, when a blue data enters the pipeline and hits the blue consumer request.
The third stage performs writes, updating table and buffers according to the mapping decision.

There are a few trade-offs making the \VLRD design simpler or more complex:
\begin{inparaenum}[1)]
\item The multiple buffer partitions 
decouple the address mapping pipeline and bus I/O,
so a burst of packets can be buffered first then fed into
the pipeline, otherwise the \VLRD just accepts one packet 
per clock cycle;
\item \LL is chosen over a bitvector to deal with the sparse buffer entry usage,
that is not only due to the consideration of FIFO property,
but also because the authors feel \LL is more scalable for large {\VLRD}s.
\end{inparaenum}
\update{Additional trade-offs are discussed in \S~\ref{subsub:endpoint}}.

%% file: routingtablediagram.tex
\begin{table*}[!b]
    \vspace{-.2in}
    \caption{Address mapping  pipeline actions
    per cycle.
    xxx$_n$ $\triangleq$ the latch for xxx in Stage n
    }
    \label{tab:pipeline}
    \centering
    \scriptsize
    \renewcommand{\arraystretch}{1.02}
    \vspace{-2pt}
    \begin{tabular}{|@{}C{0.03\linewidth}@{}|L{0.30\linewidth}|L{0.30\linewidth}|L{0.30\linewidth}|}
        \hline
        \textbf{Cyc.} & \textbf{Stage 1} reads linkTab (\SQI $\rightarrow$ head, tail) & \textbf{Stage 2} makes mapping decision (hit/miss) & \textbf{Stage 3} updates tables and buffers \\
        \hline
        1 &
        \cellcolor{LinkBlue}%
        prodHead$_1$, consTail$_1$ $\leftarrow$ NULL, NULL
        \newline
        /*~linkTab[consBuf[1].linkId], CIHR $\leftarrow$ 2 */
        &
        &
        \\
        \hline
        2 &
        \cellcolor{LinkOrange}%
        prodHead$_1$, consTail$_1$ $\leftarrow$ NULL, NULL
        \newline
        /*~linkTab[consBuf[2].linkId], CIHR $\leftarrow$ NULL */
        &
        \cellcolor{LinkBlue}%
        miss: append to the linked list in consBuf
        \newline
        /*~because prodHead$_1$=NULL, no blue data */
        & \\
        \hline
        3 &
        \cellcolor{LinkBlue}%
        consHead$_1$, prodTail$_1$ $\leftarrow$ 1, 1 /* RAW */
        \newline
        /*~linkTab[prodBuf[1].linkId], PIHR $\leftarrow$ 2 */
        &
        \cellcolor{LinkOrange}%
        miss: append to the linked list in consBuf
        \newline
        /*~because prodHead$_1$=NULL, no orange data */
        &
        \cellcolor{LinkBlue}%
        linkTab[1].cons\{Head, Tail\} $\leftarrow$ 1, 1
        /*~linkId$_2$=1, CIHR$_2$=1,
        new consHead read by Stage 1 */
        \\
        \hline
        4 &
        \cellcolor{LinkGreen}%
        consHead$_1$, prodTail$_1$ $\leftarrow$ NULL, NULL
        /*~linkTab[prodBuf[2].linkId], PIHR $\leftarrow$ 3 */
        &
        \cellcolor{LinkBlue}%
        hit: read consBuf[1] for consTgt, nextL
        \newline
        /* consHead$_1$=1 */
        &
        \cellcolor{LinkOrange}%
        linkTab[0].cons\{Head, Tail\} $\leftarrow$ 2, 2
        \newline
        /*~linkId$_2$=0, CIHR$_2$=2 */
        \\
        \hline
        5 &
        \cellcolor{LinkBlue}%
        consHead$_1$ $\leftarrow$ NULL
        /*nextL$_2$ forwarded*/
        \newline
        prodTail$_1\leftarrow$NULL
        /*linkTab[prodBuf[3].linkId]*/
        &
        \cellcolor{LinkGreen}%
        miss: append to the linked list in prodBuf
        \newline
        /* because consHead$_1$=NULL, no green request */
        &
        \cellcolor{LinkBlue}%
        linkTab[1].consHead $\leftarrow$ NULL /*~nextL$_2$ */
        \newline
        set prodBuf[1].OUT\qquad
        POHR, POTR $\leftarrow$ 1, 1
        \\
        \hline
    \end{tabular}
    \vspace{-.4in}
\end{table*}

%% file: isa.tex
\subsection{Instruction Set Extensions}\label{sec:isa}

To allow software 
to express the role of producer/consumer explicitly, \VL adds three new 
instructions  for {\bf \SELECT, \PUSH} and {\bf \FETCH} operations.
Technically
they are ``data cache'' maintenance instructions with a \texttt{dc} nomenclature;
we simply refer to them by their named function.

\noindent\textbf{\SELECT \RT}:~\label{select}
The \SELECT identifies a specific cache
line by a \VA in the operand register \RT.
As the name suggests,
\SELECT ``selects'' a cache line addressed by \VA,
so that a follow-on \PUSH or \FETCH instruction can perform its operation on the ``selected'' cache line. Through \SELECT, the \VA of the cache 
line is translated, and the \PA gets latched into a system register (not part of context state) 
only accessible by \PUSH or \FETCH.
Similar to load-linked store-conditional (\textit{LLSC}),
where a load-link always precedes a store-conditional,
there is a dependency between a \SELECT instruction and a \PUSH or \FETCH instruction,
although \FETCH itself can be executed speculatively and out-of-order with respect to instructions other than \PUSH or \FETCH.
In the case the cache line to select has been evicted into memory, \SELECT 
generates a cache miss and brings the cache line back to L1 data cache (\LONE),
just as any 
store would, in an ``exclusive'' cache state.
On context swap or page migration, 
the latched \PA is cleared. 

\noindent\textbf{\PUSH \RS, \RT}:~\label{push}
The \PUSH instruction takes the cache line from \SELECT 
and conditionally writes it from cacheable memory to a 
\VLRD memory target \RT (provided as a \VA). This \VA
in \RT is assigned to the \VLRD by the scheme described in \S~\ref{subsub:endpoint}.
The operand register \RS receives the result 
of zero for success or nonzero upon failure of a \PUSH operation. 
On completion,
the selection of the cache line ends
(i.e., \PA in the system register set by \SELECT is zeroed).
There are a few scenarios the \PUSH 
operation could fail. First, 
a \PUSH being called without a previous
\SELECT call
results in
a non-zero value written back to \RS.
The second,
is the most expected failure case where the \VLRD has no 
buffering capacity or consumer demand which also returns a non-zero to \RS. 
A system register counting \PUSH instructions on-the-fly ensures no context swap or interrupt can occur before a \RS receives a result.
The \VLRD must make forward progress in a fixed interval,
i.e. bounded by the time it takes to get to the \VLRD, which is
approximately 14 cycles in our implementation. \PUSH is a device 
memory write on the coherence network, as such, the write is 
non-snooping and it cannot be merged with other writes.

\noindent\textbf{\FETCH \RS, \RT}:
The \FETCH has the effect of pulling data from a \VLRD memory location 
(the \VA from \RT) into the calling core's private cache
at the location specified by the paired \SELECT call.
Like \PUSH, \FETCH clears cache line selection on execution.
If data is 
available on a given \SQI (see \S~\ref{subsub:endpoint} 
for \VA to \VLRD and \SQI mapping), then the \VLRD sends a data 
injection to the user buffer location specified by \SELECT immediately. 
If data is not available, the request is conditionally 
registered with the \VLRD, conditional on buffering capacity
for requests in the \VLRD.
A successful request results in a zero
value being stored in \RS.
Once data is available for the requested
\SQI then data is conditionally injected. \FETCH sets 
a ``pushable'' bit 
within the calling core's private caches, this facilitates 
asynchronous (and speculative) conditional data injection  by the
\VLRD while ensuring data still in-use is not overwritten by the 
\VLRD.
If there is a context swap, thread migration following a \FETCH,
or the line is evicted,
the injection attempt is rejected,
because the ``pushable''
cache flag is unset before any of those scenarios occur,
and the data remain with the \VLRD.
The system register set by \SELECT is cleared by \FETCH as well.
On being scheduled
the programmer is expected to check the line to see if new data
has arrived (e.g. examine control region from \S~\ref{subsec:control}),
to re-issue the request which sets the cache tag as ``pusheable'' again.

The \ISA described adds a single bit to the cache tag array of each
private cache, and adds conditional write and push commands to support
the signalling. \VL uses an otherwise standard
coherence network with non-snooping directed data transfer, the width
of that network remains unchanged. 

%% file: lib.tex
\subsection{User-space and System Software}\label{sec:lib}


Using an existing queuing framework such as \boost 
or \ZMQ with \VL is simply a matter of mapping the 
\ISA from \S~\ref{sec:isa} corresponding 
to enqueue/dequeue semantics to
the existing software queue
application program
interface.
There are a few additional allocation constraints,
such as specific alignment requirements and \VLRD setup.
Hence, we develop a library to ease the programmer burden.
  
In Figure~\ref{fig:overallDetailed}, a user
binary starts by requesting a \SQI (equivalent to a file handle) 
at \marker{1} and \marker{2}.
At \marker{3} 
the programmer maps this \SQI into a process accessible \VA through
a system call at \marker{4}, that sets up the 
\VLRD with the \SQI at \marker{5c} and returns a mapped \VA
into user-space at
\marker{5a} and \marker{5b}.

\subsubsection{\SQI allocation \& release}\label{subsub:sqi}
\MN endpoints 
assigned to a \SQI are allowed to communicate. This is akin
to ``shared memory'' \update{Inter-Process-Communication (\textit{IPC})} with the \SQI being
analogous to a file descriptor and following similar rules
with similar supervisor/OS protections~\cite{opengroup}.
The \SQI can be used to open endpoints from user-space,
granting the calling thread access to map this \SQI channel
into its address space. Listing~\ref{code:shmopen} is what is executed at \marker{1}
of Figure~\ref{fig:overallDetailed}, resulting in the \SQI at \marker{2}. 
\SQI closing and ordering semantics are identical to those
of ``shared'' memory POSIX file handles, simplifying the
programming interface. 
\begin{listing}[b]
\vspace{-4mm}
\begin{minted}
[
style=default,
frame=lines,
framesep=.5mm,
baselinestretch=1,
bgcolor=white,
fontsize=\footnotesize
]
{c++}
const int SQI = 
    shm_open(  "queue_name", O_RDWR, 
               VL_QUEUE /** flag for queue **/)
\end{minted}
\vspace{-8mm}
\caption{Example of calling a POSIX compliant \texttt{shm\_open}
with the string handle ``queue\_name'', with a read and write
mode,
and a \texttt{VL\_QUEUE} flag that tells the
supervisor that this is to be a \VL shared memory operation.}
\label{code:shmopen}
\end{listing}

\subsubsection{Endpoint creation}\label{subsub:endpoint}
As shown in Figure~\ref{fig:overallDetailed}, once a \SQI is obtained,
the programmer must ``open'' the queue \marker{3} then 
map that descriptor to a \VA to address
the assigned \VLRD. This \SQI is mapped to a \VA using
\texttt{mmap}~\cite{opengroup} (via a kernel module wrapper at \marker{5a} and \marker{5b})
as shown in Code snippet~\ref{code:mmap}
using the addressing scheme described shortly. 

\begin{listing}[b]
\vspace{-4mm}
\begin{minted}
[
style=default,
frame=lines,
framesep=.5mm,
baselinestretch=1.0,
bgcolor=white,
fontsize=\footnotesize
]
{c++}
void *X = 
    mmap( nullptr, QPAGE_SIZE, PROT, 
          VL_QUEUE /** flag for queue **/, 
          SQI, 0x0 )
\end{minted}
\vspace{-8mm}
\caption{Example of obtaining a \VA mapping for the \SQI from
user-space. The \VA returned is to a device memory location which
maps the \VA to the \PA of the \VLRD.}
\label{code:mmap}
\end{listing}

\begin{figure}[hbt]
    \centering
    \includegraphics{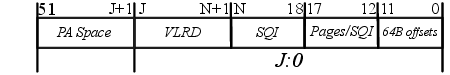}
    \caption{Device-memory \PA bit fields addressing the \VLRD
    }
    \vspace{-4mm}
    \label{fig:mapping}
\end{figure}
A user-space library can subdivide the device-memory-mapped \VA page further to 
make multiple non-overlapping (\SI{64}{\byte}-aligned) addresses for the same 
\SQI within a single address space. 
Our implementation maps a \SI{4}{\kibi \byte} page to each
page-aligned \textit{MMIO} address on the \VLRD.
A bit-vector within the
user-space wrapper around \texttt{mmap} is maintained to quickly find an unused,
\SI{64}{\byte}-aligned offset to return.
If \texttt{PROT\_WRITE} is given the library call returns
a producer page mapping, likewise if \texttt{PROT\_READ} is given, a consumer
page returned. 
Removing a user-space 
\VA mapping for an endpoint is through the \texttt{munmap} 
command~\cite{opengroup}. 

The allocated endpoint \VA from \texttt{mmap} is the means
by which \PUSH is able to target the \VLRD,
and the \PA (translated from the \VA) is the
means by which the \VLRD can determine the \SQI.
Figure~\ref{fig:mapping}, describes the bit fields
of the \PA with \VL information encoded.
A \VLRD simply takes $N:18$ as the \SQI,
while bits $J:N+1$ could distinguish different {\VLRD}s if more than one \VLRD are implemented to serve different VQs independently.
Multiple pages may be used, e.g. to map into differing address spaces,
or more than 64 endpoints are needed.
This is what bits $17:12$ used for,
allowing up to thirty two \SI{4}{\kibi\byte} pages.
This memory mapping process is repeated for the consumer endpoints.
A downside of this process is
physical address space is used,
e.g. with $1$-\textit{VLRD},
and $16$-\textit{SQI}s then $N\gets22$ and $J\gets26$ which would use up
\SI{67}{\mebi\byte} of address space (not physical memory). 
An alternative \update{addressing scheme that we 
explored adds an} address table to the \VLRD (populated on 
\texttt{mmap}) to map to arbitrary addresses,
\update{however, at the cost of an extra cycle to the
pipeline \S~\ref{sec:routing} and content addressable 
memory for the routing table}.

\subsubsection{User-space buffer creation}\label{subsub:usbc}
\VL enables both producers and consumers to use any page-aligned
cacheable memory as the user-space buffer for local endpoints
(e.g. the data source at \marker{1} from Figure~\ref{fig:highLevel}).
The memory could be obtained from any generic memory allocation functions
(e.g. \texttt{posix\_memalign}).
The capacity of these buffers can be adjusted in user-space
without impacting \VL to accommodate bursty behavior or 
non-stationary queue traffic distributions.
It is these user-space memory buffers
that are used in subsequent \textit{enqueue} and \textit{dequeue} operations (\S~\ref{subsec:control}).
The user-space buffer
for each endpoint is used as a circular buffer for sending lines to the \VLRD,
as such it will typically be kept cache-local.
Once a line from the user-space buffer is pushed to the \VLRD, it is marked as cleaned,
(e.g. reset control region as discribed in \S~\ref{subsec:control}),
so that it is
ready for follow-on \textit{enqueue} operations.


%% file: Evaluation.tex
\subsection{Experimental Methodology}\label{sec:method}

We evaluate \VLong with
7
benchmarks listed in Table~\ref{tab:benchmark}. 
To capture a wide range of communication and synchronization patterns, we 
chose to evaluate several kernels from the Ember~\cite{ember} benchmark
suite: \pingpong, \halo,
\sweep,
and \incast.
\fir is a typical digital 
signal processing 
workload that pipelines data through several filter stages.
The overhead of 
fine-grained pipelining for \fir has spawned several field programmable gate array implementations~\cite{evans1994efficient}. 
\textit{Bitonic} sorting algorithm~\cite{batcher1968sorting}
is a good candidate for fine-grained parallelization.
\update{The \pipeline~\cite{wang2016caf} benchmark emulates network package processing and has a mix of different queue patterns.}
All benchmarks are compiled using \textit{gcc-8.2.0}
and optimization level `\texttt{-O3}'.
For all experiments, affinity is set
to reduce unnecessary noise from thread migration.
A state-of-the-art software queue implementation,
Boost Lock Free Queue (\boost version 1.63)
is set as the baseline.
We also compare \VL to \ZMQ (version 4.2.1), another popular software queue implementation.

All the experiments, unless noted, are 
performed using \GEM~\cite{gem5},
with \VL hardware support implemented as extensions to an \textit{AArch64} architecture.
Table~\ref{tab:gem5_settings} summarizes key 
simulator settings.
\begin{table}[hbtp]
    \caption{Benchmarks.}
    \label{tab:benchmark}
    \centering
    \scriptsize
    \renewcommand{\arraystretch}{1.02}
    \begin{tabular}{L{0.17\linewidth}L{0.74\linewidth}}
        \textbf{Benchmark} & \textbf{Description, (\MN)$\times k$ $\triangleq$ producer:consumer $\times$ channel} \\
        \hline
        \pingpong~\cite{ember} & data back and forth between two threads (1:1)$\times 2$
        \\
        \hline
        \halo~\cite{ember} & exchange data with neighboring threads (1:1)$\times 48$
        \\
        \hline
        \sweep~\cite{ember} & data sweeps through a grid of threads corner to corner (1:1)$\times 48$
        \\
        \hline
        \incast~\cite{ember} & all threads sending data to the master thread (15:1)$\times 1$
        \\
        \hline
        \fir & data streams through \update{32-}stage FIR filter (1:1)$\times \update{31}$
        \\
        \hline
        \bitonic~\cite{batcher1968sorting} & bitonic sort with varying number of threads (1:N)$\times 1 + $(M:1)$\times 1$
        \\
        \hline
        \update{\pipeline~\cite{wang2016caf}} & \update{4-stage pipeline with middle stages multi-threaded (1:4)$\times 1 + $(4:4)$\times 1 + $(4:1)$\times 1 + $ (1:1)$\times 1$}
        \\
    \end{tabular}
    \vspace{-2mm}
\end{table}

\begin{table}[hbtp]
\vspace{-2mm}
    \caption{\GEM Simulator Hardware Configuration.}
    \label{tab:gem5_settings}
    \centering
    \scriptsize
    \renewcommand{\arraystretch}{1.02}
    \begin{tabular}{L{0.14\linewidth}|L{0.77\linewidth}}
        \textbf{Cores} & 16 $\times$ \textit{AArch64} OoO CPU @ \SI{2}{\giga\hertz} \\
        \hline
        \multirow{2}{*}{\textbf{Caches}} & \SI{32}{\kibi \byte} private 2-way L1D, \SI{48}{\kibi\byte} private 3-way L1I \\
         & \SI{1}{\mebi \byte} shared 16-way mostly-inclusive L2 \\
        \hline
        \textbf{Memory} & \SI{8}{\gibi \byte} \SI{2400}{\mega\hertz} DDR4 \\
        \hline
        \update{\textbf{VLRD}} & \update{64 entries per \prodBuf, \consBuf, and \linkTab (about \SI{5}{\kibi \byte} in total)} \\
    \end{tabular}
\vspace{-2mm}
\end{table}
\begin{table}[hbt]
\vspace{-2mm}
\caption{Hardware platforms}
\label{table:hwplatform}
    \centering
    \scriptsize
    \renewcommand{\arraystretch}{1.1}
\begin{tabular}{@{}L{0.02\linewidth}@{}C{0.62\linewidth}@{}L{0.26\linewidth}L{0.05\linewidth}}
\multicolumn{2}{@{}l}{\textbf{Platform}\qquad\qquad\qquad\textbf{Processor}} &  \textbf{Memory} & \textbf{OS}      \\
\hline
1                 & \begin{tabular}[c]{@{}l@{}}AMD 2990WX 32-Core @ \SI{3.2}{\giga\hertz}\end{tabular}                   & \SI{128}{\gibi\byte} DDR4-3200 & \multirow{2}{0.04\linewidth}{Linux 5.4}   \\
\cline{0-2}
2                 & \begin{tabular}[c]{@{}l@{}}Intel E5-2690v3 12-Core dual socket @ \SI{2.6}{\giga\hertz}\end{tabular}  & \SI{64}{\gibi\byte} DDR4-2133 &   \\
\end{tabular}
\vspace{-5mm}
\end{table}

\subsection{Results and Analysis}\label{sec:results}

\begin{figure}[!hbtp]
  \vspace{-3mm}
  \begin{minipage}{\linewidth}
    \centering
    \subfloat[{\color{red}Execution time (\SI{}{\nano \second}) normalized to \boost}]{
    \label{fig:speedup_allbmk}
    \includegraphics[
    width=\linewidth,
    natwidth=6.08in,
    natheight=2.49in,
    clip=true]{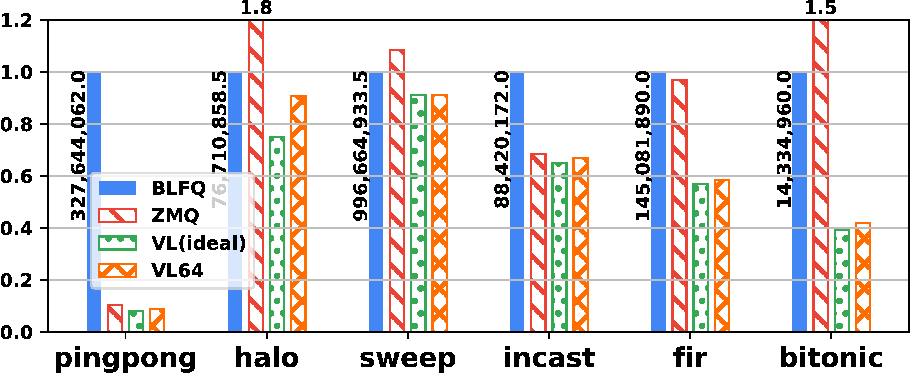}
    }
  \end{minipage}
  \begin{minipage}{\linewidth}
    \centering
    \subfloat[Snoop traffic normalized to \boost]{
    \label{fig:snoop_allbmk}
    \includegraphics[
    width=\linewidth,
    natwidth=6.08in,
    natheight=2.49in,
    clip=true]{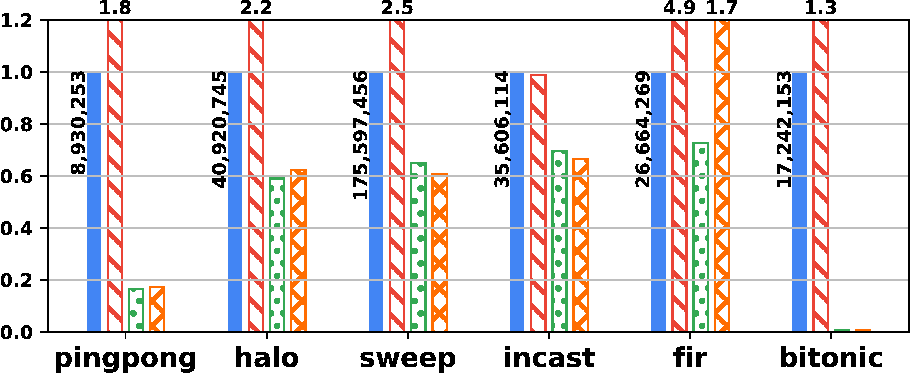}
    }
  \end{minipage}
  \begin{minipage}{\linewidth}
    \centering
    \subfloat[Memory transactions normalized to \boost]{
    \label{fig:memtrans_allbmk}
    \includegraphics[
    width=\linewidth,
    natwidth=6.08in,
    natheight=2.49in,
    clip=true]{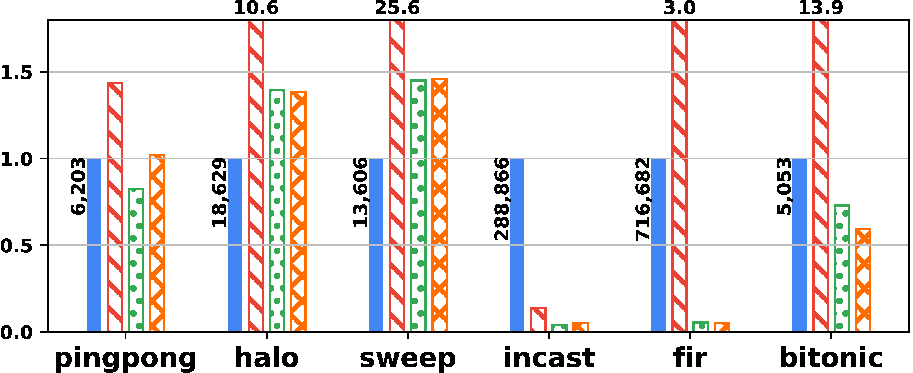}
    }
  \end{minipage}
  \caption{Comparison between different queues.}
  \label{fig:allbmk_comp}
  \vspace{-4.5mm}
\end{figure}

In Figure~\ref{fig:allbmk_comp},
we compare \VL with two state-of-the-art software queues,
\boost as baseline and \ZMQ.
In addition to this, we add {\VL}(ideal) which
has \update{\textbf{infinity}} queue capacity and
\update{\textbf{zero-latency}} cache line transfers
\update{in order to show that those hardware limitations do not put much overhead on \VL}.
Each \VL run is given with $64$ buffer entries,
and denoted as {\VL}64. 

In Figure~\ref{fig:speedup_allbmk} we see that \VL
is on average $2.09\times$ faster than software solutions,
ranging from $11.36\times$ faster for \pingpong to
$1.10\times$
faster for
\sweep.
\ZMQ falls
somewhere in between on all benchmarks, though notably 
being slower on \halo and \bitonic, which both favor low-latency small message traffic. 
However, on \incast and \fir,
\boost builds up a long queue
spilling to memory
(many more memory transactions in Figure~\ref{fig:memtrans_allbmk}),
\ZMQ and \VL both have a back-pressure mechanism so get better performance.
Figure~\ref{fig:snoop_allbmk} shows the relative magnitude of snoop 
transactions initiated per benchmark and with queue schemes.
\VL has fewer snoops than either of the two software queues (\boost and \ZMQ). 
The only exception \fir has two threads per core creating many context switches, which lead to more frequent failures for VLRD's attempts to deliver cache lines.
Software queues suffer from more snoop transactions due to cache 
coherence (as discussed in \S~\ref{sec:motiv}),
while \VLong reduces the snoop traffic to a minimum 
as it reduces the cache coherent state shared 
between communicating threads. 
Figure~\ref{fig:memtrans_allbmk} compares the amount of memory transactions between queues.
Overall, \VL has the fewest memory transactions among the queuing schemes.
\VL and \ZMQ are significantly lower on \incast and \fir with the help of the back-pressure mechanism.
On \pingpong and \bitonic,
\VL also achieves about 20\% reduction compared to \boost,
while \ZMQ has more memory transactions.
\VL has more memory transactions on \halo, and \sweep,
because the benchmarks double buffer the 
communication channels and not all the buffers are managed 
by our provided queuing libraries, but by the application. 

\begin{figure}[b]
    \vspace{-4mm}
    \centering
    \includegraphics[
        page=1,
        natwidth=5.90in,
        natheight=2.54in,
        clip=true,
        width=0.99\linewidth
    ]{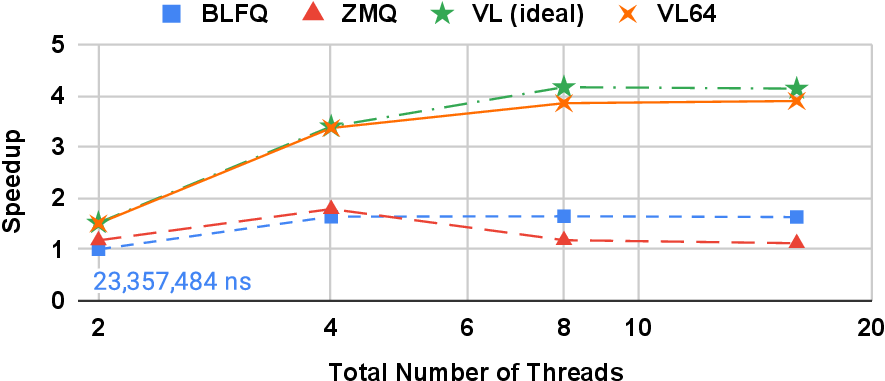}
    \caption{The scalability of \bitonic.}
    \label{fig:bitonic_scaling}
    \vspace{-2mm}
\end{figure}

\begin{figure}[!hbt]
    \centering
    \includegraphics[
        page=1,
        natwidth=6.71in,
        natheight=3.17in,
        clip=true,
        width=0.89\linewidth
    ]{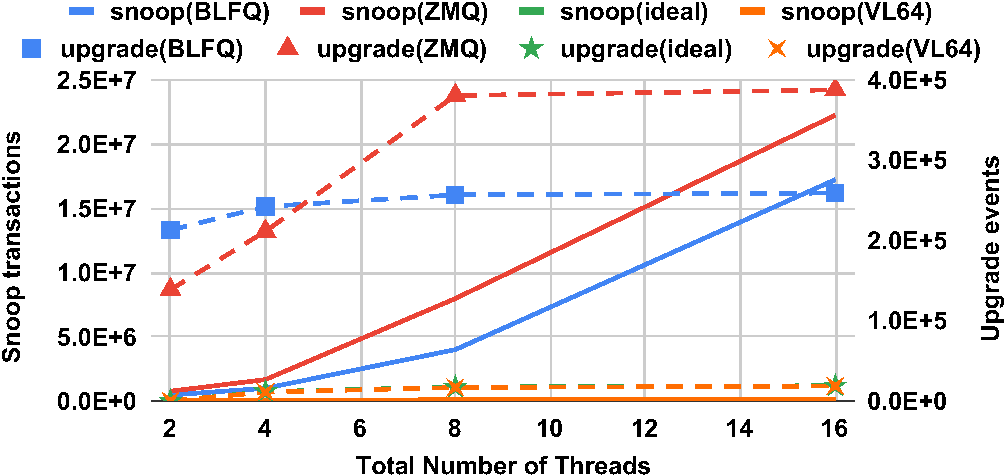}
    \caption{Snoop and upgrade events as \bitonic scales.}
    \label{fig:upgrade_snoop}
    \vspace{-5mm}
\end{figure}

\noindent\textbf{Scalability:}
\textit{Bitonic} has a fixed workload divided 
among a varying number of worker threads. 
Figure~\ref{fig:bitonic_scaling} presents 
the scalability of \bitonic with various queue implementations
as the number of worker threads are changed
(1, 3, 7, and 15 worker threads plus one master thread dispatching tasks to worker threads).
Initially, \ZMQ performs better than \boost with small numbers of threads (i.e., 2, 4),
but {\ZMQ}'s performance drops after 8 threads.
The high overhead to maintain cache coherence (as shown in Figure~\ref{fig:upgrade_snoop}) degrades the performance of \ZMQ.
Because \boost does \CAS operations, it scales slightly better than 
\ZMQ, however, neither scale as well as \VL. 
\boost stops scaling by 4 threads. 
In contrast, \VL is still able to gain speedup moving from 4 threads to 8 threads.
At 8 threads, the computation part of the single master thread dominates the execution time and become the bottleneck; 
that is why none of the queuing mechanisms can help any more.
In Figure~\ref{fig:upgrade_snoop},
we present one big difference
between \VL and the other software queues at a microarchitecture level,
to better understand why they scale differently.
Both the \boost and \ZMQ software implementations have more cache line upgrade events than \VL,
and the rate of snoop traffic synchronization goes more rapidly. 
\VL has very few upgrades and snoops,
therefore it is able to scale better than \boost and \ZMQ (see Figure~\ref{fig:lockbehavior}). 

\begin{figure}[!b]
    \vspace{-4mm}
    \centering
    \includegraphics[
        page=1,
        natwidth=6.01in,
        natheight=2.51in,
        clip=true,
        width=0.89\linewidth
    ]{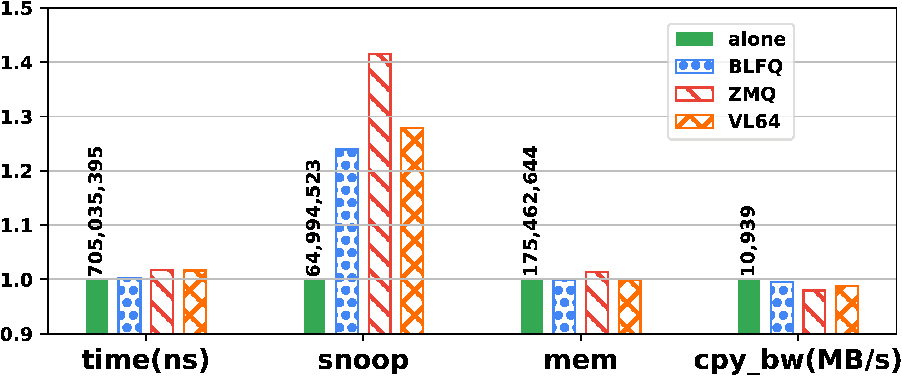}
    \caption{\update{Performance impact of message channels
    on a memory intensive application (\stream~\cite{mccalpin1995memory}).
    Each bar represents, \stream (alone), \stream  with \pingpong (\boost/\ZMQ/\VL).
    }}
    \label{fig:snoop_interence}
    \vspace{-2mm}
\end{figure}

\update{
\noindent\textbf{Coherence traffic interference:}
\VL channels use the coherence network for moving data between cores.
This could impact the coherence traffic patterns and hurt the 
performance of other applications that do not use \VL.
To study the impact of \VL on coherence traffic,
we ran the \stream benchmark~\cite{mccalpin1995memory} concurrently with \pingpong using 
each queue implementation (\boost, \ZMQ, \VL).
\stream was chosen as it is known to stress the 
memory hierarchy. 
Figure~\ref{fig:snoop_interence} shows that the
execution time for each queue implementation (\boost, \ZMQ, \VL)
varied by $2\%$ or less when compared to \stream executing alone. 
The other three bar groups report the system snoop and memory traffic.
The snoop traffic introduced by \VL is comparable to that of \boost, 
and significantly lower than that of \ZMQ.}

\update{
\noindent\textbf{Area estimation:}
We developed RTL code for the \VLRD (control logic + buffers),
synthesized it using the Synopsys Design Compiler with the FreePDK \SI{45}{\nano \meter} 
library~\cite{openpdk}, and scaled the design to \SI{16}{\nano \meter} using ~\cite{tech_scaling}
for comparison purposes.
The resulting \VLRD area is \SI{0.142}{\square \milli \meter} for buffers
and \SI{0.155}{\square \milli \meter} in total including control logic.
To put this into perspective, an Arm A-72 core at 16FF is
\SI{\sim 1.15}{\square \mm }~\cite{A72microarch_report};
our design is $13\%$ of the single-core area, however, each \VLRD
is meant to serve $N$ cores. A $16$-core Arm A-$72$ configuration (like our simulation), 
excluding \LTWO caches and wire overhead, would be approximately 
\SI{18.4}{\square \milli \meter}.
Based on this estimation,
our \VLRD
shared by 16 cores, would occupy less than $1\%$ of overall \SOC area
(adding caches and wire area would only improve this ratio). 

}

\begin{figure}[!hbt]
    \centering
    \includegraphics[
        page=1,
        natwidth=6.08in,
        natheight=1.50in,
        clip=true,
        width=0.89\linewidth
    ]{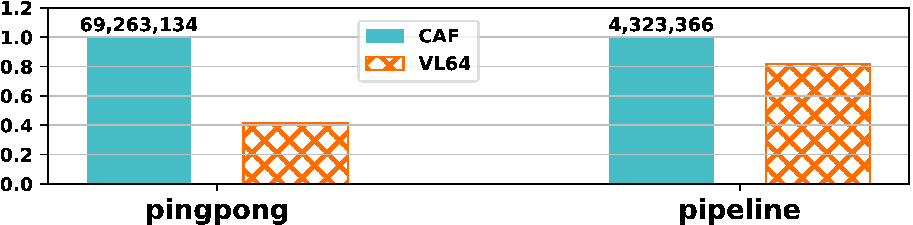}
    \caption{\update{Performance comparison between CAF~\cite{wang2016caf} and \VL.}}
    \label{fig:vscaftime}
    \vspace{-2.5mm}
\end{figure}
\update{
\noindent\textbf{Comparison with CAF:}
CAF~\cite{wang2016caf} is a state-of-the-art hardware queue proposal similar to \VL
with a couple of differences:
i. CAF divides buffers between queues and applies advanced credit management for QoS,
while buffers in \VLRD are shared by all queues;
ii. The enqueue/dequeue operations in CAF transfer 64-bit values
between registers and Queue Management Device,
whereas \VL exploits cache lines as local buffer and as such
lowers the frequency of performing relatively
more costly data movement through the cache hierarchy.
We compare \VL with CAF on two benchmarks used in CAF paper, \pingpong and \pipeline:
\pingpong passes data through the queue,
while \pipeline uses the queue for pointers to \SI{2}{\kibi \byte} network packet payloads.
As shown in Figure~\ref{fig:vscaftime},
\VL achieves $2.40\times$ speedup over CAF on \pingpong, and
$1.22\times$ speedup on \pipeline.
}

%% file: Related.tex
Software \IPC ranges from POSIX standard
\IPC (e.g. \texttt{mkfifo}~\cite{opengroup}) to user-space 
libraries such as the \boost~\cite{boost} (a more complete survey 
can be found in~\cite{herlihy2020art}). 
Instead of focusing on 
improving the algorithms, \VL focuses on hardware-software
codesign, arriving at a solution that
combines the flexibility of software with hardware acceleration. 

\IPC is closely related, even synonymous, to message-passing (including MPI),
data-flow, stream-processing, and many other topics.
MPI, generically, is a topic of constant research, recent works~\cite{jose2010unifying,hjelm2016evaluation,pritchard2020getting} in 
particular focus on reducing 
core-to-core latency. These works expose very low-level tuning knobs to the programmer
assuming the programmer can better tune an application. Our work on 
the other hand focuses on maintaining the same programming
semantics expected by even novice parallel programmers while reducing 
Dataflow processing, is closely related to systolic array processing, 
stream processing, and Coarse-grained Reconfigurable
Array (\textit{CGRA}) processing. Loosely, the aforementioned topics
are collected together as they all aim to allow maximum  exploitation 
of spatial communication patterns, 
allowing each \PE to send data directly to down-stream dataflow 
targets~\cite{dennis1980data,arvind1982u}. 
Dataflow connections forming communications links are often direct 
register-to-register transfers mediated by a common bus
(e.g.,~\cite{grafe1989epsilon,papadopoulos1990monsoon,noakes1993j} and many others 
summarized by~\cite{lee1993issues,hurson2007dataflow}).
Systolic arrays are also a form of dataflow, although with a fixed
spatial communication pattern, e.g. ~\cite{pomerleau1988neural,dally2003merrimac,jouppi2018motivation}.
Closely related to the above are \textit{CGRA}s~\cite{gray1993configurable}. 
Each work differs slightly in the amount of reconfiguration
permitted, from the least flexible systolic array to the most
flexible \textit{CGRA}. Unlike these, \VL can exploit 
spatial locality of data streams while having dynamic software configurable
connectivity. 


Modern core-to-core communication concepts, occupy a spectrum
from direct memory transfer instructions to various hardware-software 
schemes. Network processing cores such as TILE64~\cite{TILE64_ISSCC08},
DSP-like processors such as the IBM Cell~\cite{chen2007cell}, 
and the Freescale DPAA~\cite{qoriq2012primer}
provide channel operators or primitives to send data from \PE-to-\PE. 
Works such as HAQu~\cite{lee2011haqu}
which uses two new \update{structures} per-core, including a \textit{Queue
Local Table}, whereas with \VL, the logic is simpler and located
within the interconnect, enabling any type of device to theoretically
connect and use \VL. HAQu decentrialized head and tail pointers 
to each core, by doing so it made \MN communications difficult to implement. 
CAF~\cite{wang2016caf} and Intel DLB~\cite{IntelDLB}, went in a different direction, centralizing the 
queue management enabling \MN. \VL goes in a different direction
entirely, focusing on minimalism in implementation while enabling \MN
and decentralized head and tail pointers.
Other works, such as~\cite{du2019xpc} focus on specific use-cases
in Android, whereas \VL intends to be more generic. 

%% file: Conclusion.tex
In this paper, we presented \VLong (\VL) an inter-process
communication (\IPC) mechanism for fine-grained multi-threaded applications.
\VL is immune to cache contention for synchronization,
provides back-pressure to reduce memory spills,
and achieves low-latency cache injection by directly stashing the line into consumer \LONE cache.
This novel cross-core synchronization mechanism is similar to software queue mechanisms in flexibility but has the performance and efficiency of hardware solutions.
Our full-system \GEM simulation illustrated
that we can obtain a $2.09\times$ speedup and $61\%$ average reduction
in memory traffic over state-of-the-art software
solutions across a variety of communications patterns and benchmarks.

%% file: main.bbl
\begin{thebibliography}{10}
\providecommand{\url}[1]{#1}
\csname url@samestyle\endcsname
\providecommand{\newblock}{\relax}
\providecommand{\bibinfo}[2]{#2}
\providecommand{\BIBentrySTDinterwordspacing}{\spaceskip=0pt\relax}
\providecommand{\BIBentryALTinterwordstretchfactor}{4}
\providecommand{\BIBentryALTinterwordspacing}{\spaceskip=\fontdimen2\font plus
\BIBentryALTinterwordstretchfactor\fontdimen3\font minus
  \fontdimen4\font\relax}
\providecommand{\BIBforeignlanguage}[2]{{%
\expandafter\ifx\csname l@#1\endcsname\relax
\typeout{** WARNING: IEEEtran.bst: No hyphenation pattern has been}%
\typeout{** loaded for the language `#1'. Using the pattern for}%
\typeout{** the default language instead.}%
\else
\language=\csname l@#1\endcsname
\fi
#2}}
\providecommand{\BIBdecl}{\relax}
\BIBdecl

\bibitem{dennard1974design}
R.~H. Dennard, F.~H. Gaensslen \emph{et~al.}, ``Design of ion-implanted
  {MOSFET}'s with very small physical dimensions,'' \emph{IEEE Journal of
  Solid-State Circuits}, vol.~9, no.~5, pp. 256--268, 1974.

\bibitem{kish2002end}
L.~B. Kish, ``End of moore's law: thermal (noise) death of integration in micro
  and nano electronics,'' \emph{Physics Letters A}, vol. 305, no. 3-4, pp.
  144--149, 2002.

\bibitem{vetter2018extreme}
J.~S. Vetter, R.~Brightwell \emph{et~al.}, ``Extreme heterogeneity
  2018-productive computational science in the era of extreme heterogeneity:
  Report for doe ascr workshop on extreme heterogeneity,'' USDOE Office of
  Science (SC), Washington, DC (United States), Tech. Rep., 2018.

\bibitem{kleen2009linux}
A.~Kleen, ``Linux multi-core scalability,'' in \emph{Proceedings of Linux
  Kongress}, 2009.

\bibitem{dijkstra1965solution}
E.~W. Dijkstra, ``A solution of a problem in concurrent programming control,''
  September 1965.

\bibitem{lamport1979make}
L.~Lamport, ``How to make a multiprocessor computer that correctly executes
  multiprocess progranm,'' \emph{IEEE transactions on computers}, no.~9, pp.
  690--691, 1979.

\bibitem{arvind1988assessing}
D.~C. Arvind and G.~Maa, ``Assessing the benefits of fine-grain parallelism in
  dataflow programs,'' \emph{International Journal of High-performance
  Computing Applications}, vol.~2, no.~3, 1988.

\bibitem{ampere}
``Ampere reveals “quicksilver” altra lineup, 128-core “mystique”
  kicker,'' \url{https://bit.ly/2Hiqj3D}, accessed: 2020-07-21.

\bibitem{pasetto2012performance}
D.~Pasetto, M.~Meneghin \emph{et~al.}, ``Performance evaluation of interthread
  communication mechanisms on multicore/multithreaded architectures,'' in
  \emph{Proceedings of the 21st international symposium on High-Performance
  Parallel and Distributed Computing}, 2012, pp. 131--132.

\bibitem{akkan2013hpc}
H.~Akkan, M.~Lang \emph{et~al.}, ``Hpc runtime support for fast and power
  efficient locking and synchronization,'' in \emph{2013 IEEE International
  Conference on Cluster Computing}.\hskip 1em plus 0.5em minus 0.4em\relax
  IEEE, 2013, pp. 1--7.

\bibitem{grafe1989epsilon}
V.~G. Grafe, G.~S. Davidson \emph{et~al.}, ``The epsilon dataflow processor,''
  \emph{ACM SIGARCH Computer Architecture News}, vol.~17, no.~3, pp. 36--45,
  1989.

\bibitem{papadopoulos1990monsoon}
G.~M. Papadopoulos and D.~E. Culler, ``Monsoon: an explicit token-store
  architecture,'' in \emph{ACM SIGARCH Computer Architecture News}, vol.~18,
  no. 2SI.\hskip 1em plus 0.5em minus 0.4em\relax ACM, 1990, pp. 82--91.

\bibitem{noakes1993j}
M.~D. Noakes, D.~A. Wallach \emph{et~al.}, ``The j-machine multicomputer: An
  architectural evaluation,'' \emph{ACM SIGARCH Computer Architecture News},
  vol.~21, no.~2, pp. 224--235, 1993.

\bibitem{TILE64_ISSCC08}
S.~{Bell}, B.~{Edwards} \emph{et~al.}, ``Tile64 - processor: A 64-core soc with
  mesh interconnect,'' in \emph{2008 IEEE International Solid-State Circuits
  Conference - Digest of Technical Papers}, Feb 2008, pp. 88--598.

\bibitem{chen2007cell}
T.~Chen, R.~Raghavan \emph{et~al.}, ``Cell broadband engine architecture and
  its first implementation—a performance view,'' \emph{IBM Journal of
  Research and Development}, vol.~51, no.~5, pp. 559--572, 2007.

\bibitem{qoriq2012primer}
D.~QorIQ, ``Primer for software architecture,'' Technical report, Freescale
  Semiconductor Inc, Tech. Rep., 2012.

\bibitem{ladan2004optimistic}
E.~Ladan-Mozes and N.~Shavit, ``An optimistic approach to lock-free fifo
  queues,'' in \emph{International Symposium on Distributed Computing}.\hskip
  1em plus 0.5em minus 0.4em\relax Springer, 2004, pp. 117--131.

\bibitem{michel2018packet}
O.~Michel, J.~Sonchack \emph{et~al.}, ``Packet-level analytics in software
  without compromises,'' in \emph{10th {USENIX} Workshop on Hot Topics in Cloud
  Computing (HotCloud 18)}, 2018.

\bibitem{boost}
``Class template queue,'' \url{https://bit.ly/37hAMHJ}, accessed: 2020-08-19.

\bibitem{hintjens2010zeromq}
P.~Hintjens, ``Zeromq: the guide,'' \emph{URL http://zeromq. org}, 2010.

\bibitem{lockhammer}
``\textit{lockhammer},'' \url{https://bit.ly/3kbvz7N}, accessed: 2020-07-21.

\bibitem{scott2013shared}
M.~L. Scott, ``Shared-memory synchronization,'' \emph{Synthesis Lectures on
  Computer Architecture}, vol.~8, no.~2, pp. 1--221, 2013.

\bibitem{martin2012chip}
M.~M. Martin, M.~D. Hill \emph{et~al.}, ``Why on-chip cache coherence is here
  to stay,'' \emph{Communications of the ACM}, vol.~55, no.~7, pp. 78--89,
  2012.

\bibitem{harchol2013performance}
M.~Harchol-Balter, \emph{Performance modeling and design of computer systems:
  queueing theory in action}.\hskip 1em plus 0.5em minus 0.4em\relax Cambridge
  University Press, 2013.

\bibitem{bertsimas1995distributional}
D.~Bertsimas and D.~Nakazato, ``The distributional little's law and its
  applications,'' \emph{Operations Research}, vol.~43, no.~2, pp. 298--310,
  1995.

\bibitem{jiang2020cimat}
H.~Jiang, X.~Peng \emph{et~al.}, ``Cimat: A compute-in-memory architecture for
  on-chip training based on transpose sram arrays,'' \emph{IEEE Transactions on
  Computers}, 2020.

\bibitem{leon2011cache}
E.~A. Le{\'o}n, R.~Riesen \emph{et~al.}, ``Cache injection for parallel
  applications,'' in \emph{Proceedings of the 20th international symposium on
  High performance distributed computing}, 2011, pp. 15--26.

\bibitem{flynn1997amba}
A.~AMBA, ``Amba-5 architecture specification,'' \url{https://bit.ly/356Sjjf},
  2020, accessed: 2020-10-13.

\bibitem{farshin2020reexamining}
A.~Farshin, A.~Roozbeh \emph{et~al.}, ``Reexamining direct cache access to
  optimize i/o intensive applications for multi-hundred-gigabit networks,'' in
  \emph{2020 {USENIX} Annual Technical Conference}, 2020, pp. 673--689.

\bibitem{huang2019comparative}
Z.~Huang, ``A comparative study on the performance isolation of virtualization
  technologies,'' Ph.D. dissertation, Arizona State University, 2019.

\bibitem{rao2013optimizing}
J.~Rao, K.~Wang \emph{et~al.}, ``Optimizing virtual machine scheduling in numa
  multicore systems,'' in \emph{2013 IEEE 19th International Symposium on High
  Performance Computer Architecture}.\hskip 1em plus 0.5em minus 0.4em\relax
  IEEE, 2013, pp. 306--317.

\bibitem{opengroup}
The open group base specifications issue 7, 2018 edition ieee std 1003.1-2017
  (revision of ieee std 1003.1-2008). \url{https://bit.ly/2Hfww0w}. Accessed
  October 2020.

\bibitem{virtio}
Virtual {I/O} {D}evice ({VIRTIO}) {V}ersion 1.1. \url{https://bit.ly/3jaEqWf}.
  Accessed October 2019.

\bibitem{RevereAMU_ARM2019}
\BIBentryALTinterwordspacing
\emph{{Revere-AMU System Architecture}}, Arm Limited, September 2019. [Online].
  Available: \url{https://bit.ly/3kajJuQ}
\BIBentrySTDinterwordspacing

\bibitem{ember}
``\textit{Ember} communication pattern library,'' \url{https://bit.ly/3k9egUV},
  accessed: 2020-10-13.

\bibitem{evans1994efficient}
J.~B. Evans, ``Efficient fir filter architectures suitable for fpga
  implementation,'' \emph{IEEE Transactions on Circuits and Systems II: Analog
  and Digital Signal Processing}, vol.~41, no.~7, pp. 490--493, 1994.

\bibitem{batcher1968sorting}
K.~E. Batcher, ``Sorting networks and their applications,'' in
  \emph{Proceedings of the April 30--May 2, 1968, spring joint computer
  conference}, 1968, pp. 307--314.

\bibitem{wang2016caf}
Y.~Wang, R.~Wang \emph{et~al.}, ``Caf: Core to core communication acceleration
  framework,'' in \emph{2016 International Conference on Parallel Architecture
  and Compilation Techniques (PACT)}.\hskip 1em plus 0.5em minus 0.4em\relax
  IEEE, 2016, pp. 351--362.

\bibitem{gem5}
\BIBentryALTinterwordspacing
N.~Binkert, B.~Beckmann \emph{et~al.}, ``The gem5 simulator,'' \emph{SIGARCH
  Comput. Archit. News}, vol.~39, no.~2, p. 1–7, Aug. 2011. [Online].
  Available: \url{https://doi.org/10.1145/2024716.2024718}
\BIBentrySTDinterwordspacing

\bibitem{mccalpin1995memory}
J.~D. McCalpin \emph{et~al.}, ``Memory bandwidth and machine balance in current
  high performance computers,'' \emph{IEEE computer society technical committee
  on computer architecture newsletter}, vol.~2, no. 19--25, 1995.

\bibitem{openpdk}
J.~E. {Stine}, I.~{Castellanos} \emph{et~al.}, ``Freepdk: An open-source
  variation-aware design kit,'' in \emph{2007 IEEE International Conference on
  Microelectronic Systems Education (MSE'07)}, 2007, pp. 173--174.

\bibitem{tech_scaling}
\BIBentryALTinterwordspacing
A.~Stillmaker and B.~Baas, ``Scaling equations for the accurate prediction of
  cmos device performance from 180nm to 7nm,'' \emph{Integration}, vol.~58, pp.
  74 -- 81, 2017. [Online]. Available:
  \url{http://www.sciencedirect.com/science/article/pii/S0167926017300755}
\BIBentrySTDinterwordspacing

\bibitem{A72microarch_report}
``\textit{Inside ARM’s Cortex-A72 microarchitecture},''
  \url{https://bit.ly/3sf0a9h}, accessed: 2021-01-09.

\bibitem{herlihy2020art}
M.~Herlihy, N.~Shavit \emph{et~al.}, \emph{The art of multiprocessor
  programming}.\hskip 1em plus 0.5em minus 0.4em\relax Newnes, 2020.

\bibitem{jose2010unifying}
J.~Jose, M.~Luo \emph{et~al.}, ``Unifying upc and mpi runtimes: experience with
  mvapich,'' in \emph{Proceedings of the Fourth Conference on Partitioned
  Global Address Space Programming Model}, 2010, pp. 1--10.

\bibitem{hjelm2016evaluation}
N.~Hjelm, ``An evaluation of the one-sided performance in open mpi,'' in
  \emph{Proceedings of the 23rd European MPI Users' Group Meeting}, 2016, pp.
  184--187.

\bibitem{pritchard2020getting}
H.~P. Pritchard~Jr, T.~Naughton \emph{et~al.}, ``Getting it right with open
  mpi: Best practices for deployment and tuning of open mpi,'' Los Alamos
  National Lab.(LANL), Los Alamos, NM (US), Tech. Rep., 2020.

\bibitem{dennis1980data}
J.~B. Dennis, ``Data flow supercomputers,'' \emph{Computer}, no.~11, pp.
  48--56, 1980.

\bibitem{arvind1982u}
A.~Arvind and K.~P. Gostelow, ``The u-interpreter,'' \emph{Computer}, no.~2,
  pp. 42--49, 1982.

\bibitem{lee1993issues}
B.~Lee and A.~R. Hurson, ``Issues in dataflow computing,'' in \emph{Advances in
  computers}.\hskip 1em plus 0.5em minus 0.4em\relax Elsevier, 1993, vol.~37,
  pp. 285--333.

\bibitem{hurson2007dataflow}
A.~R. Hurson and K.~M. Kavi, ``Dataflow computers: Their history and future,''
  \emph{Wiley Encyclopedia of Computer Science and Engineering}, 2007.

\bibitem{pomerleau1988neural}
D.~A. Pomerleau, G.~L. Gusciora \emph{et~al.}, ``Neural network simulation at
  warp speed: How we got 17 million connections per second,'' CMU, Tech. Rep.,
  1988.

\bibitem{dally2003merrimac}
W.~J. Dally, F.~Labonte \emph{et~al.}, ``Merrimac: Supercomputing with
  streams,'' in \emph{Proceedings of the 2003 ACM/IEEE conference on
  Supercomputing}.\hskip 1em plus 0.5em minus 0.4em\relax IEEE, 2003, pp.
  35--35.

\bibitem{jouppi2018motivation}
N.~Jouppi, C.~Young \emph{et~al.}, ``Motivation for and evaluation of the first
  tensor processing unit,'' \emph{IEEE Micro}, vol.~38, no.~3, pp. 10--19,
  2018.

\bibitem{gray1993configurable}
J.~Gray and T.~Kean, ``Configurable hardware: a new paradigm for computation,''
  in \emph{Proceedings, 10th Cultech. Conference on VLSI}, 1993, pp. 279--295.

\bibitem{lee2011haqu}
S.~Lee, D.~Tiwari \emph{et~al.}, ``Haqu: Hardware-accelerated queueing for
  fine-grained threading on a chip multiprocessor,'' in \emph{2011 IEEE 17th
  International Symposium on High Performance Computer Architecture}.\hskip 1em
  plus 0.5em minus 0.4em\relax IEEE, 2011, pp. 99--110.

\bibitem{IntelDLB}
``\textit{Queue Management and Load Balancing on
  Intel\textsuperscript{\textregistered} Architecture},''
  \url{https://intel.ly/3hY0Zy8}, accessed: 2021-01-09.

\bibitem{du2019xpc}
D.~Du, Z.~Hua \emph{et~al.}, ``{XPC}: architectural support for secure and
  efficient cross process call,'' in \emph{Proceedings of the 46th
  International Symposium on Computer Architecture}, 2019, pp. 671--684.

\end{thebibliography}
